\documentclass[article]{aa}  
\usepackage{graphicx, natbib} 
\usepackage{txfonts}
\newcommand{\sax}{{\it BeppoSAX}}

\newcommand{\integral}{{\it INTEGRAL}}

\newcommand{\source}{SAX~J1712.6--3739}

\begin{document}
\titlerunning{The broad band X-ray spectra of the LMXB SAX J1712.6-3739}
   \title{
SAX~J1712.6-3739: a persistent hard X-ray source as monitored with {\it INTEGRAL}{\thanks
{{\it INTEGRAL\/} is an ESA project with instruments and Science Data Centre funded by ESA member states (especially the PI countries: Denmark, France, Germany, Italy, Switzerland, Spain), Czech Republic and Poland, and with the participation of Russia and the USA.
}} 
}

   \subtitle{}

   \author{Mariateresa Fiocchi\inst{1}, Angela Bazzano\inst{1}, Pietro Ubertini\inst{1}, Giovanni De Cesare
          \inst{1}\inst{2}\inst{3}
          }

   \offprints{}

   \institute{\inst{1}Istituto di Astrofisica Spaziale e Fisica Cosmica di Roma (INAF). Via Fosso del Cavaliere 100, Roma, I-00133, Italy\\
\inst{2}Dipartimento di Astronomia, Universita' degli Studi di Bologna, Via
 Ranzani 1, I40127 Bologna, Italy\\
\inst{3}Centre d'Etude Spatiale des Rayonnements, CNRS/UPS, B.P. 4346, 31028
 Toulouse Cedex 4, France\\
              \email{mariateresa.fiocchi@iasf-roma.inaf.it}
             }

   \date{Received ..........., accepted ............}
 
  \abstract
   {The X-ray source \source\/ is a very weak Low Mass X-ray Binary discovered 
in 1999 with \sax\/ and
 located in the Galactic Center. This region has been deeply investigated
 by the \integral\/ satellite with an unprecedented exposure time,
 giving us an unique opportunity
 to study the hard X-ray behavior also for weak objects.
   }
   {We analysed all available \integral\/ public and private Key-Program observations
 with the main aim of studying
 the long-term behavior of this Galactic bulge X-ray burster.
   }
   {The spectral results are based on the systematic analysis of all \integral\/
 observations covering the source position performed
 between February 2003 and October 2006. 
\source\/ did not shows any flux variation along this period as well 
as compared to previous \sax\/ observation. Hence, to better constrain the 
physical parameters   
we combined both instrument data.
}
    {Long \integral\/ monitoring reveals that this X-ray burster
 is a weak persistent source,
displaying a X-ray spectrum extended to high energy and spending most of the time in a low luminosity hard state.
The broad-band spectrum is well modeled with a simple Comptonized model
with a seed photons temperature of $\sim$ 0.5 keV and an electron 
temperature of $\sim$24 keV.
 The low mass accretion rate ($\sim$ 2$\times$10$^{-10}$ M$_{sun}$/yr), 
the long bursts
 recurrence time, the small sizes of the region emitting the
 seed photons consisting with the
 inner disk radius and the high luminosity ratio
 in the 40--100 keV and 20--40 keV band, are all features common to the Ultra Compact source class.
   }
{  We report, for the first time, on the X-ray behavior of this source: observations with unprecedent wide energy band 
and sensitivity revealed this X-ray
 burster is a persistent source with an hard spectrum extending up to energies $>$ 100 keV.
}


   \maketitle
%

\section{Introduction}
\source\/ was discovered in 1999 
during the monitoring campaign of the Galactic Centre region performed 
by the \sax\//WFC (Cocchi et al. 1999 and in 't Zand et al. 1999). 
This instrument observed a type-I X-ray burst with exponential decays on August 1999, 
identifying 
the compact object as a weakly magnetized neutron star in a low mass X-ray 
binary system with a derived distance of 7 kpc (Cocchi et al. 2001).
The source was followed-up by the \sax\//instruments to investigate 
the spectral properties: 
\source\/ shows a spectrum extended up to 
about 60 keV, with an intensity of $\sim$ 6 mCrab, both in the MECS (1--10 keV) and 
in the PDS (15--60 keV).
The spectrum was fitted by an absorbed power law ($\Gamma\sim$2.2)
and a column density N$_H$$\sim$2$\times$10$^{22}$cm$^{-2}$ or by a Comptonised 
emission model with electron temperature of $\sim$ 50 keV and a lower 
column density N$_H$$\sim$1$\times$10$^{22}$cm$^{-2}$ (Cocchi et al. 2001).
The ROSAT source J171237.1-373834 
is just 0\farcm6 from the \sax\/ position with a flux of 1.6 mCrab in 
the energy range 0.1--2.4 keV 
(in 't Zand et al. 1999). 
No optical counterpart has been identified yet within the 13\arcsec\
(1$\sigma$) ROSAT error circle radius.
\source, monitored during the PCA/RXTE bulge scan program,
is continuously active, apparently in two states: a slowly changing
state, and a quicker one (in 't Zand et al. 2007). 
It was included in the IBIS/ISGRI soft Gamma-Ray survey catalog (Bird et al. 2007) 
as a transient X-ray buster
with fluxes corresponding to $4.7\pm0.1$ mCrab and $4.0\pm0.2$ mCrab 
in the energy bands 20--40 keV and 40--100 keV, respectively.
Recently, Chelovekov et al. (2006) reported
two further burst detections with \integral\//IBIS in 2003-2004, 
both with the same peak flux.

\section{Observations and Data Analysis}
\sax\/ observed the source on September 1999 with an exposure of 21 ksec. 
LECS, MECS and PDS event files and spectra,
available from the ASI Scientific Data Center,
were generated by means of the Supervised Standard Science Analysis
\cite{fio99}.
Both LECS and MECS spectra were accumulated in circular regions
of 8' radius.
The PDS spectra were obtained
with the background rejection method based on fixed rise time thresholds.
Publicly available matrices were used for all the instruments.
The cross-calibration constant
values were taken in agreement with the indications given in Fiore et al. 1999.
Fits are performed in the following energy band: 0.5--3.0 keV for LECS, 1.5--10.0
keV for MECS and 15--70 keV for PDS.\\
The \integral\/ (Winkler et al. 2003) observations 
are divided into uninterrupted 2000-s intervals, the so-called science windows (SCWs).
Wideband spectra (from 5 to 200 keV) of the source are obtained using data from the
two high-energy
instruments JEM-X \citep{lun03} and IBIS \citep{ube03}.
Data were processed using the Off-line Scientific Analysis
(OSA v5.1 and v6.0 for IBIS and JEM-X, respectively)
software released by the \integral\/ Scientific Data Centre.
While IBIS  provide a very large FOV ($>30\degr$), JEM-X has
a narrower FOV ($>10\degr$), 
thus providing only a partial overlap with the high-energy detectors.
Data from the Fully Coded field of view only have been used for both instrument.
Two X-ray bursts detected with \integral\//IBIS in 2003-2004 reported by
Chelovekov et al. (2006) are not included in our data set, being
the source out of our selected field of view.
IBIS light curves and spectra are extracted for each individual SCW.
JEM-X spectrum was extracted from mosaic image at the position of the \source. 

\section{Spectral Analysis of the persistent emission}
During the \sax\/ observations the source intensity was $\sim$ 6 mCrab
both in the MECS (2-10 keV) and PDS (15-60 keV) instruments. 
The \source\/ flux level after discovery was near the limiting sensitivity of the WFC instrument
(Jager et al. 1997) and
for this reason this object was classified as a transient source (Cocchi et al. 1999). 
\integral\/ long monitoring, at the limiting sensitivity 
of $\sim$ 1 mCrab at 10 $\sigma$ level on an exposure of $\sim$10 Ms,
shows that \source\/ is a weak persistent source,
with a time averaged intensity of $6.1\pm0.3$~mCrab and $6.7\pm0.7$~mCrab,
for IBIS (17-30 keV) and JEM-X (4-10 keV) instruments, respectively.
The \source\/ luminosity do not significantly vary
from one observation to another, so that we can take advantages in order:
1) to create an high quality IBIS and JEM-X average spectra 
with long exposure time (685 ks and 132 ks, respectively) and 2)  
to build up a wide band spectrum with no gap from 0.5 to 200 keV,
combining
LECS, MECS, PDS, IBIS and JEM-X data.\\
The whole
data set was fitted with several physical models,
using XSPEC v.\ 11.3.1.

We fitted the spectrum in the energy band 0.5--200 keV
with a simple cutoff powerlaw model but this resulted in a poor fit to the 
data with a reduced chi square $\chi^2_\nu\sim2.7$ (135 d.o.f.). 
Similar result ($\chi^2_\nu\sim3.1$ with 135 d.o.f.) was obtained fitting this data set with the
Comptonization model {\scriptsize{COMPST}} 
(Sunyaev and Titarchuk 1980).

   \begin{figure*}[t]
   \centering
\includegraphics[width=6cm, angle=-90]{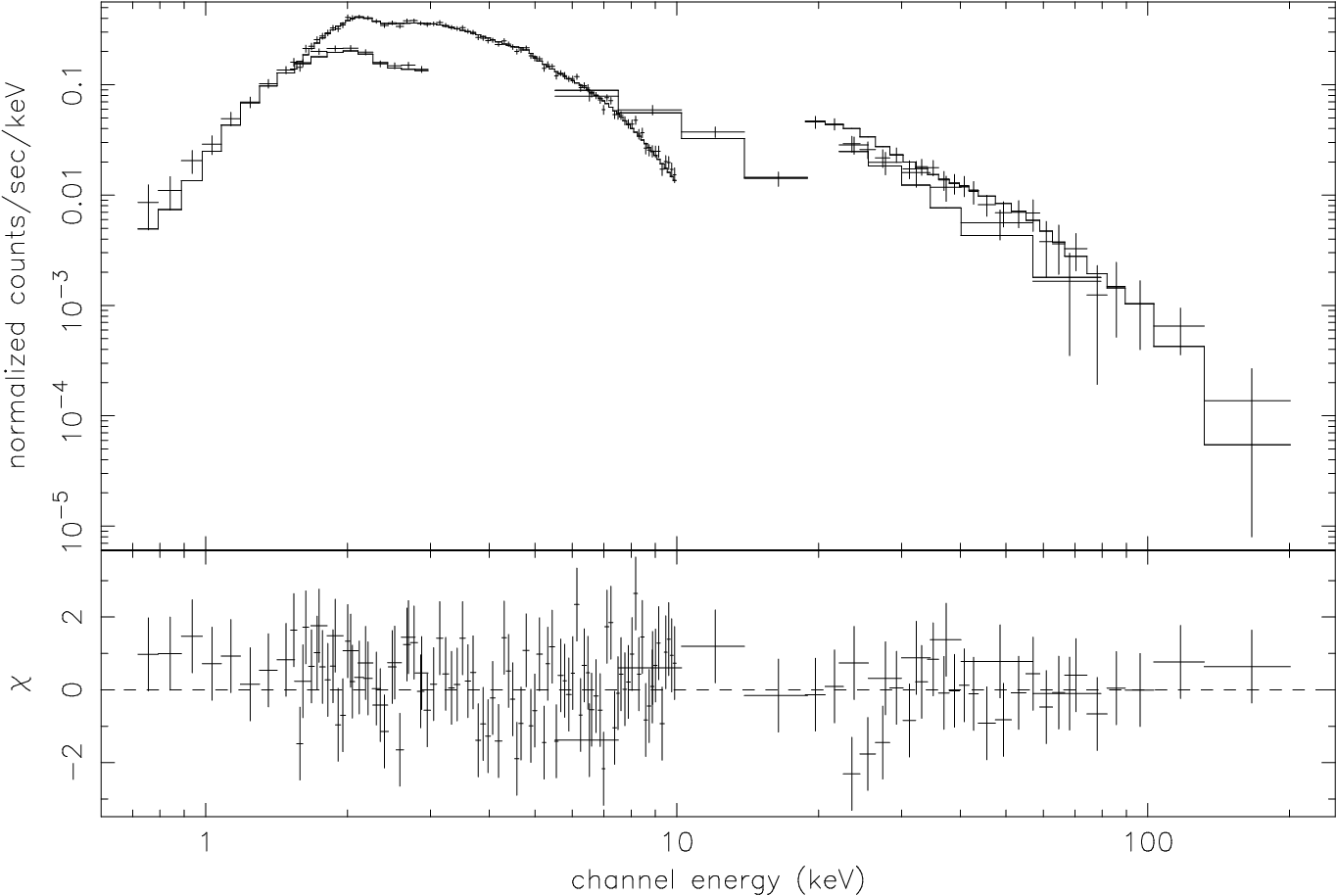}
   \includegraphics[width=6cm, angle=-90]{comptt.cps}
      \caption{\integral\/ and \sax\/ data and residuals respect to the 
best model 
consisting of a simple comptonized {\scriptsize{COMPTT}} model ({\em left})
and respective energy spectrum ({\em right}).
Light blue points are \sax\/ data and red points are \integral\/.
             }
         \label{fig1}
   \end{figure*}

   \begin{figure*}[t]
   \centering
\includegraphics[width=6cm, angle=-90]{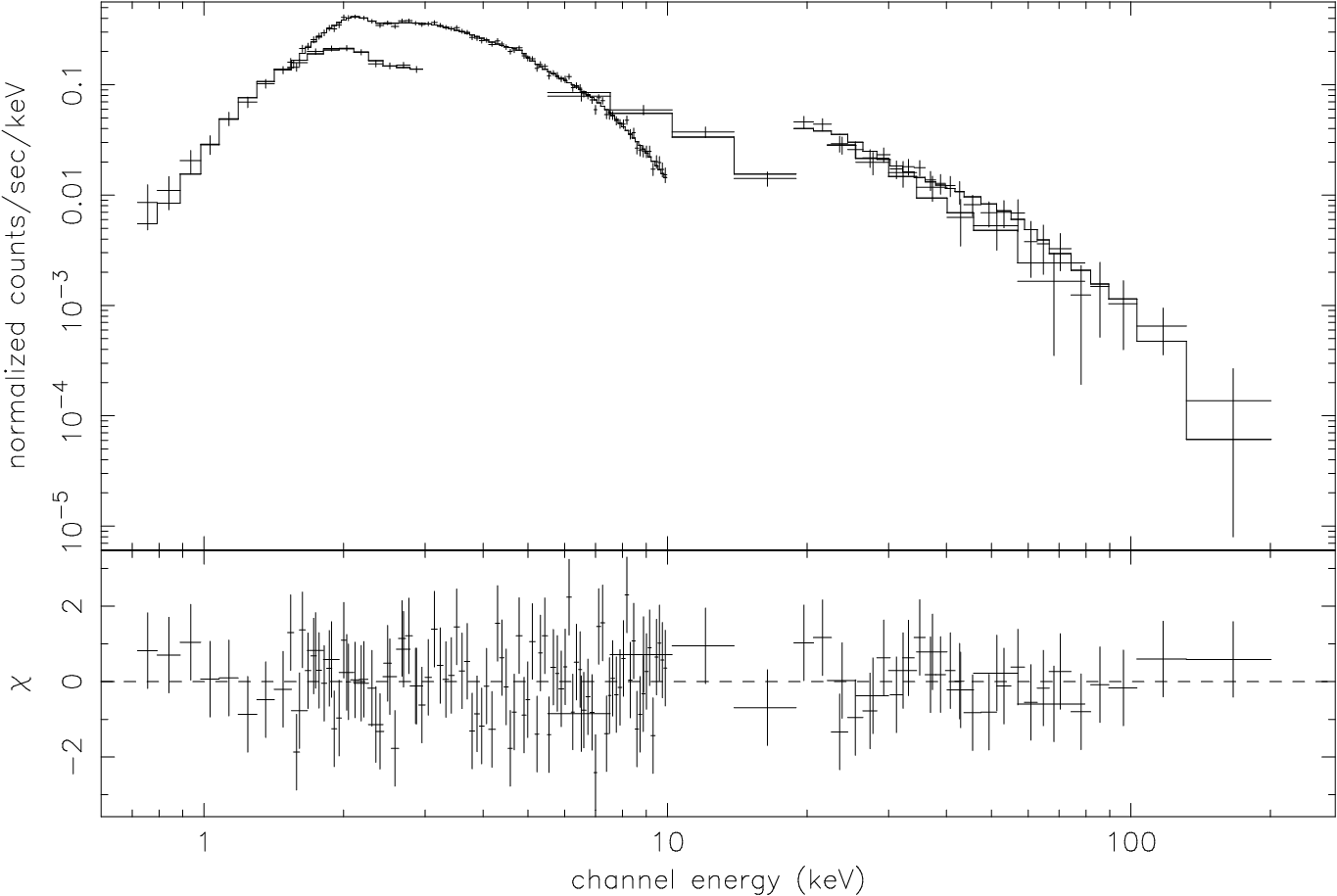}
   \includegraphics[width=6cm, angle=-90]{plot_diskbb2.cps}
      \caption{\integral\/ and \sax\/ data and residuals respect to the
best model
consisting of a multicolor black body and Comptonized {\scriptsize{COMPTT}} model ({\em left})
and respective energy spectrum ({\em right}).
Light blue points are \sax\/ data and red points are \integral\/.
The contributions of the two component are shown separately:
orange and green are multi color black body and Comptonization component,
respectively, and their sum is plotted in magenta.
             }
         \label{fig3}
   \end{figure*}

   \begin{figure*}[t]
   \centering
\includegraphics[width=6cm, angle=-90]{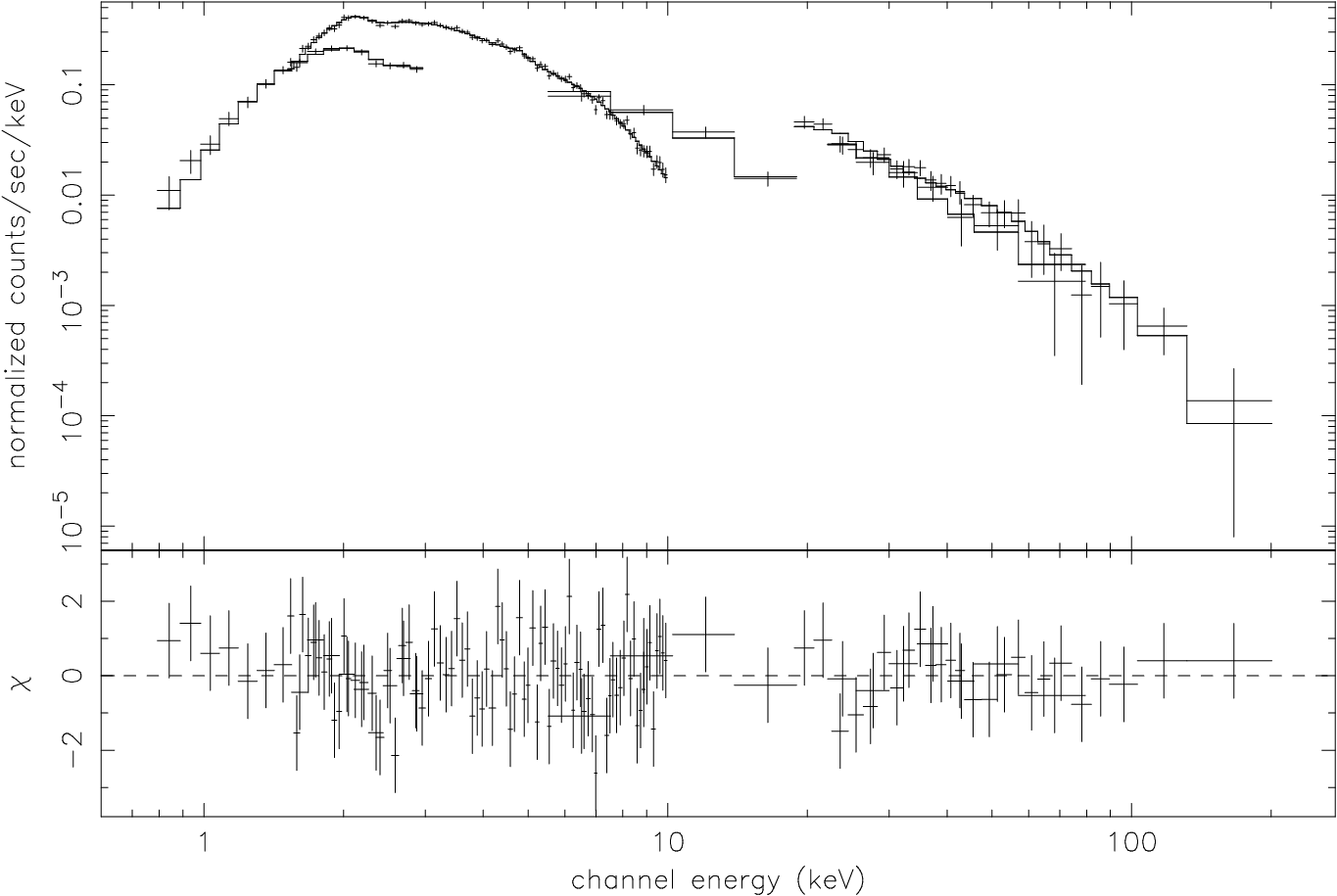}
   \includegraphics[width=6cm, angle=-90]{plot_bb2.cps}
      \caption{\integral\/ and \sax\/ data and residuals respect to the
best model
consisting of a simple black body and a Comptonized {\scriptsize{COMPTT}} model ({\em left})
and respective energy spectrum ({\em right}).
Light blue points are \sax\/ data and red points are \integral\/.
The contributions of the two component are shown separately:
orange and green are black body and Comptonization component,
respectively, and their sum is plotted in magenta.
             }
         \label{fig2}
   \end{figure*}

For both models, the strongest residuals are at the energy below few keV, 
indicating that a more complex model with a soft component 
is need to fit these data. This is agreement with the picture 
formulated in the past years according to which the X-ray spectra of the low mass 
X-ray binaries were generally described as the sum of a soft and hard component.
These in turn are interpreted in the framework of two models: 
the Eastern Model (Mitsuda et al. 1984) consisting of a blackbody
together with a Comptonized component and the
Western model (White et al. 1984), consisting of a multicolor disk blackbody
plus a Comptonized component.
For the soft component we used the black body
component in the formulation of Mitsuda et al. (1984) with all parameters free
or the multicolor disk blackbody component in the formulation of
Makishima et al. \cite{m:86}, which
assume the gravitational energy
released by the accreting material is locally dissipated into
blackbody radiation, that the accretion flow is continuous throughout
the disk, and finally the effects of electron scattering are negligible.
The two parameters of this model are free:
${\rm r_{in}({\cos}i)^{0.5}}$ where ${\rm r_{in}}$ is the innermost
radius of the disk, {\em i} is the inclination angle of the disk and
${\rm kT_{in}}$ the blackbody effective temperature at ${\rm r_{in}}$.
Adding a black body component improves the fit significantly, reducing
$\chi^2/dof$ from 366/135 to 138/133 for the cutoff power law model and 
from 414/135 to 160/133 for the {\scriptsize{COMPST}} model.
We also have tried to fit this data set with a 
model consisting of a only one {\scriptsize{COMPTT}} (Titarchuk 1994) component,
assuming a spherical geometry
for the Comptonizing region. 
This replaces
the {\scriptsize{COMPST}} Comptonization model in the sense that the theory
is extended to include relativistic effects and the seed photon energy can
be in the soft X-ray range, while in the older models the 
seed photons are cool, i.e. in the UV band.
We left the following parameters free: the temperature of the Comptonizing
electrons kT${\rm _e}$, the plasma optical depth 
${\rm \tau_p}$ and the input  
temperature of the soft photon Wien distribution kT${\rm _0}$.
This model gives a good $\chi^2/dof$ of 142/134 and also take care of
the photons deficit below few keV with respect to 
the cutoff powerlaw or {\scriptsize{COMPST}} models.
The physicals parameters obtained using this model are reported 
in Table~\ref{fitsax}.
Adding a soft X-ray component to the {\scriptsize{COMPTT}} component 
reduces $\chi^{2}/$d.o.f.\ from 142/134 to 119/132 for the black body component
and from 142/134 to 120/132 for multicolor disk blackbody component, 
with a low F-test chance probabilities of $8.6\times 10^{-6}$ and 
$1.5\times 10^{-5}$, respectively.
{{These fit results of the broad band spectrum using the models with 
two emission components
are reported
in Table~\ref{fitsax}. }}

The column density {N$_{\rm H}$} towards the source
was left free and its value measured by the LECS and MECS instrument
is always close to the galactic column density
($N_{\rm H}~galactic~=~1.34\times10^{22}~cm^{-2}$,
estimated from the 21 cm measurement of Dickey \& Lockman 1990).
Figures 1, 2 and 3 show the counts spectra with residuals respect to the 
best fits 
and the photon spectra of the \source\/
using the models described above. These figures clearly illustrate 
the importance of the broad band coverage obtained combining \integral\/ and \sax\/
data, which allowed us to well constrain the physical parameters
and clearly shows the source has a constant luminosity from $\sim 3$ up to 20 keV.

\section{Discussion}

\integral\/ observations reveal for the first time that this 
X-ray burster is a persistent source, displaying a very hard X-ray spectrum 
and spending most of the time in a low luminosity and hard state.
A broad-band spectrum obtained combining the \integral\/ together with
the \sax\/ data 
displays properties which fits well into the classification of low luminosity 
($\sim$0.01  $ L_{\rm Edd}$) weakly magnetized neutron stars with 
"hard spectra'' (Barret et al. 2001): a broad-band spectrum extending up to 100 keV,
 dominated by a Comptonized component, with a optical depth of the 
Comptonizing corona of $\sim$3, seed photons temperature of $\sim$0.5 keV, 
an electron temperature of $\sim$24 keV. A few LMXBs appear to spend most 
of the time in this "hard state'' (Di Salvo \& Stella 2002), 
e.g. 4U 0614+091 (Piraino et al. 1999) or 4U1850-087 (Sidoli et al. 2006).

The emission spectrum could be also fitted with a two component
model consisting of a blackbody, or multi color blackbody component which 
represents emission from an optically thick accretion disk or from the
neutron star surface, together with a  Comptonized component
which is interpreted as emission
from a hot inner disk region or a
boundary layer between the disk and the neutron star.
We have estimate 
a black body radius of $\sim 5$ km for the black body component 
and the inner radius of the disk
$< 11 km$ for the multi disk component.
Although the data statistics does not allow us to distinguish 
between the two physical models, both values of the radius give an 
indication of the system compactness.

We have also estimated the sizes of the emitting seed photons region, 
assuming their emission as blackbody with temperature T$_0$ 
and radius R$_{seed}$ following in 't Zand et al. (1999). 
We obtain $R_{\rm seed}\simeq 8.3\,{\rm km}\, 
(L_{\rm C}/10^{37}\,{\rm erg\, s}^{-1})^{1/2} (1+y)^{-1/2} (kT_0/1\,{\rm keV})^{-2}$, 
where $L_{\rm C}$ is the luminosity of the Comptonization component. 
Here, we have estimated the amplification factor of the seed luminosity 
by the Comptonization as $(1+y)$, where $y$ is the Comptonization parameter, 
$y=4kT_{\rm e}\max(\tau,\tau^2)/m_{\rm e} c^2$. 
The obtained values are $\sim$ 1 km and $\sim$ 10 km for the model with the 
blackbody component and with multicolor disk blackbody component, respectively.
The parameters obtained using the latter model  
allows for a more reasonable 
interpretation.

\begin{table*}
\vspace{1cm}
\caption{Results of the \source\/ fit, using different models:
a simple comptonized model, a blackbody plus
a Comptonized component and
a multicolor disk blackbody
plus a Comptonized component.
Uncertainties are at the 90\% confidence level for a single parameter variation.}
\label{fitsax}
\vspace{0.3cm}
\centering
\begin{tabular}{ccccccccc}
\hline\hline
&& & & & &  &&\\
Model 1&{N$_{\rm H}$} &...&...&  {T$_{\rm o}$}& {T$_{\rm e}$} &$\tau$ & $n_{\rm COMPTT}$ & $\chi^2$/d.o.f \\
&($10^{22}~cm^{-2}$)&...&... &($keV$) &(keV)&...&($10^{-3}$)&...\\
&& & & & & & &\\
$comptt$&$1.3\pm0.1$&...&... &$0.47\pm0.02$ &$24^{+16}_{-10}$  &$2.7\pm1.3$ &$2.0^{+1.2}_{-0.8}$&142/134\\
&& & & & & & &\\
\hline                                                                              && & & & & & &\\
Model 2&{N$_{\rm H}$} & {T$_{\rm in}$}& ${\rm r_{in}({\cos}i)^{0.5}}$& {T$_{\rm o}$}& {T$_{\rm e}$} &$\tau$ & $n_{\rm COMPTT}$ & $\chi^2$/d.o.f \\
&($10^{22}~cm^{-2}$) &($keV$)&(km)&($keV$) &(keV)&...&($10^{-3}$)&...\\
&& & & & & & &\\
$diskbb+comptt$&$1.4\pm0.3$&$0.7\pm0.3$  &$<11$ &$0.5\pm0.2$ &$24^{+22}_{-9}$  &$3\pm2$ &$1.6^{+1.6}_{-0.8}$&120/132\\
&& & & & & & &\\
\hline
&& & & & & & &\\
Model 3 &{N$_{\rm H}$} & {T$_{\rm bb}$}& {R$_{\rm bb}$}& {T$_{\rm o}$}& {T$_{\rm e}$}
&$\tau$ & $n_{\rm COMPTT}$ & $\chi^2$/d.o.f \\
&($10^{22}~cm^{-2}$) &($keV$)&(km)&($keV$) &(keV)&...&($10^{-3}$)&...\\
&& & & & & & &\\
$bbody+comptt$&$1.2\pm0.2$&$0.6\pm0.1$&$5\pm1$&$1.2\pm0.2$ &$34^{+54}_{-14}$ &$2.0\pm1.5$ &$0.5^{+0.6}_{-0.2}$&119/132\\
&& & & & & & &\\
\hline
\hline
\end{tabular}\\
\end{table*}

These results show that the radius of 
the region providing the seed photons 
for the Comptonization emission is consistent with the inner radius of the
accretion disk, showing that 
the seed photons are bounded in small region
spatially coincident with the inner part of the disk. 
Moreover the seed photons temperature $T_0$ is consistent with 
the inner disk temperature $T_{in}$.
By imposing $T_0$=$T_{in}$, the fit values correspond to a disk temperature of
$0.57\pm0.08$ keV with a $\chi^2/dof$ of 121/133.  
This behavior suggests to identify the inner 
disk as the region providing the seed photons of the Comptonization emission.
This result 
was observed in most of ultra compact X-ray 
source: 4U~1850-087, 4U~1820-303 and 4U~0513-40 
(Sidoli et al. 2001), XTE J1751-305 (Gierli\'nski \&  Poutanen 2005), 
XTE J1807-294 (Falanga et al. 2005).

Assuming the persistent flux represents the average mass accretion rate and  
an accretion efficiency of $\eta = 0.2$ 
(corresponding, e.g., to $M_{\rm NS}=1.4 {\rm M}_{\sun}$ 
and $R_{\rm NS}=10$ km) and using multi color disk model luminosity, $L_{0.1-200 keV}=1.6\times10^{36} erg s^{-1}$,
we get
$\dot{M}\simeq 2.8 \times 10^{-10} {\rm M}_{\sun}$ yr$^{-1}$ for \source.
This low mass accretion rate is consistent with the hard spectrum 
of \source\/, according to our present understanding on the X-ray bursters 
(for review see van der Klis 2006).
Moreover, the burst recurrence time is dependent on the mass accretion rate on
the neutron star: the faster new fuel is provided from the donor star, the shorter is the 
burst recurrence time. 
Long \integral\/ monitoring shows clearly that the frequency of bursts in this sources 
is low, as previously reported by in 't Zand et al. (2007), who estimated 
a recurrence time of 
345--6507 hr.

In 't Zand et al. (2007) have selected all 31 persistent X-ray bursters as reported in the 
\integral\/ survey (Bird et al. 2006) and demonstrated
that the low mass accretion rates are accompanied by the hard X-ray spectra
of the persistent emission. Almost all the Ultra Compact X-ray binaries have 
the highest values of the (40--100)keV/(20--40)keV hardness ratio.
\source\/ was not included as supposed to have a transient nature.
We now know it is a persistent and weak source,
showing a high   
ratio L$_{(40-100)keV}$/L$_{(20-40)keV}$$\sim$0.90,
in agreement with the value of 0.85$\pm$0.06 as derivated from recent data in Bird et al. 2007.     
Finally, the low mass accretion rate, the long bursts
 recurrence time, the small sizes of the region emitting the
 seed photons consisting with the
 inner disk radius and the high luminosity ratio
 in the 40--100 keV and 20--40 keV 
support the idea of In 't Zand et al. (2007)
that this source could be a candidate Ultra Compact Binary.

\begin{acknowledgements}
We are grateful to Steven N. Shore for his interest in and
support of this work.
We acknowledge the ASI financial/programmatic support via contracts ASI-IR
I/023/05/0.
A special thank to M. Federici for supervising
the \integral\/ data archive and to
in 't Zand for useful scientific discussions and suggestions.
\end{acknowledgements}


\begin{thebibliography}{}
\bibitem[]{ba01}
 Barret, D. 2001, in Adv. Space Res., 28, 307
\bibitem[]{bi06}
Bird A. J., et al., 2006, \apj, 636, 765
\bibitem[]{bi07}
Bird A. J., et al., 2007, \apj~S, in press
\bibitem[]{Chelovekov06} Chelovekov, I.V., Grebenev, S.A., \& Sunyaev, R.A.,
        2006, AstL, 32, 456 (astro-ph/0605638)
\bibitem[]{coc99} Cocchi, M., Natalucci, L., in 't Zand, J.J.M., et al. 1999,
        IAUC 7247
\bibitem[]{coc01} Cocchi, M., Bazzano, A., Natalucci, L., Heise, J., \&
        in 't Zand, J.J.M. 2001, Mem. S.A.It., 72, 757
\bibitem[]{di90}
Dickey \& Lockman, 1990, Ann. Rev. Astron. Astrophys. 28, 2
\bibitem[]{di03}
 Di Salvo, T., \& Stella, L. 2002, proceedings of the XXII Moriond Astrophysics Meeting, The Gamma-Ray Universe (Les Arcs, March 9-16, 2002), ed. A. Goldwurm, D. Neumann, \& J. Tran Thanh Van (Vietnam: The Gioi Publishers) [arXiv:astro-ph/0207219]
\bibitem[]{faa05}
Falanga, M., Bonnet-Bidaud, J. M., Poutanen, J., Farinelli, R., Martocchia, A., Goldoni P., Qu J. L., Kuiper, L., Goldwurm, A., A\&A, 2005, 436, 647
\bibitem[Fiore et al. 1999]{fio99}
Fiore, F., Guainazzi, M., \& Grandi, P. 1999,
Cookbook for BeppoSAX NFI Spectral Analysis (www.asdc.asi.it/bepposax/software/cookbook)
\bibitem[]{gi05}
Gierli\'nski, M., \&  Poutanen, J., 2005, MNRAS, 359, 1261
\bibitem[]{zan99} in 't Zand, J.J.M., Heise, J., Bazzano, A., Cocchi, M., \&
        Smith, M.J.S. 1999, IAUC 7243
\bibitem{zan02} in 't Zand, J.J.M., Markwardt, C.B., Bazzano, A.,
        et al. 2002, A\&A, 390, 597
\bibitem[]{zan07}
in 't Zand, J.J.M., Jonker, P. G., \& Markwardt C. B., 2007, A\&A, accepted.
\bibitem[]{ja97}
Jager et al., 1997, A\&AS, 125, 557
\bibitem[Lund et al. 2003]{lun03}
Lund, N., et al., 2003, A\&A, 411, L231
\bibitem[1986]{m:86}
Makishima K., Maejima Y., Mitsuda K., et al., 1986, ApJ 285, 712
\bibitem[]{mi86}
Mitsuda, K., Inoue, H., Koyama, K., Makishima, K., Matsuoka, M., Ogawara, Y.,
Suzuki, K., Tanaka, Y., Shibazaki, N., Hirano, T., 1984, PASJ, 36, 741
\bibitem[]{pi}
Piraino, S., Santangelo, A., Ford, E. C., Kaaret, P.,
1999, A\&A, 349, 77
\bibitem{sid99} Sidoli, L., Parmar, A.N., Oosterbroek, T., Stella, L.,
        Verbunt, F., Masetti, N., \& Dal Fiume, D. 2001, A\&A, 368, 451
\bibitem{sid06}
L. Sidoli, A. Paizis, A. Bazzano, S. Mereghetti, 2006, A\&A, 460, 229.
\bibitem{sun80}
Sunyaev and Titarchuk, 1980, A\&A, 86, 121
\bibitem[1994]{ti:94}
Titarchuk L., 1994, ApJ 434 570
\bibitem[Ubertini et al. 2003]{ube03} Ubertini, R., Lebrun, F., di Cocco, G. et al.
2003,
A\&A,  411, 131
\bibitem[van]{van}
Van der Klis, M., 2006, in "Compact Stellar X-ray Sources", eds. W.H.G. Lewin and
M. van der Klis, CUP (Cambrige, UK).
\bibitem[White et al. 1988]{whi88}
White N. E., Stella L. \& Parmar A. N., 1988, \apj, 324, 363
\bibitem[Winkler et al. 2003]{wi03}
Winkler, C., Gehrels, N., Schönfelder, V., Roques, J.-P., Strong, A. W., Wunderer, C
., Ubertini, P., 2003, A\&A, 411, 349
\end{thebibliography}
\end{document}